\documentstyle[preprint,revtex]{aps}
\begin{document}
\begin{title}
Self Organization and a Dynamical Transition in Traffic Flow Models
\end{title}
\author
{Ofer Biham and A. Alan Middleton }
\begin{instit}
Department of Physics,
Syracuse University,
Syracuse, NY 13244
\end{instit}
\author {Dov Levine}
\begin{instit}

Department of Physics,
Technion, Israel Institute of Technology,
Haifa, 32000, Israel
\end{instit}

\begin{abstract}
A simple model that describes traffic flow in two dimensions is
studied. A sharp {\it jamming transition } is found that separates
between the low density dynamical phase in which all cars move at maximal
speed and the high density jammed phase  in which they are all stuck.
Self organization  effects in
both phases are studied and discussed.
\end{abstract}

\pacs{02.70.+d, 05.70.Ln, 64.60.Cn, 89.40.+k}

Traffic problems have been studied  extensively in recent years
in order to help in the design of transportation infrastructure
and to optimize the allocation of resources.
Traffic simulations, based on various hydrodynamic models, have
provided much insight and are in good agreement with experiments for simple
systems such as a freeway, a tunnel or a single junction
\cite{traffic}.
However, the simulation of traffic flow in a whole city
is a formidable task as it involves
many degrees of freedom such as local densities and speeds.
The availability of powerful
supercomputers is likely to  make these simulations feasible in the
near future,
but models that are simpler and more flexible than hydrodynamic models
will be needed in order to achieve this task.

Cellular automaton (CA) models
\cite{wolf}
are increasingly used in
simulations of complex physical systems such as fluid dynamics
\cite{lattgas},
driven diffusive systems
\cite{schnitt},
sand piles
\cite{bak}
and chemical reactions
\cite{bzreac}.
In some of these systems the cellular
automaton models provide only some general qualitative features of the
system while in other cases useful quantitative information can be obtained.
For some problems involving complex geometries, such as
simulations of fluid dynamics in porous media,
cellular automata are found to be superior
to other methods.

In this paper we present three variants of  a simple  cellular automaton
model that describes traffic flow in two dimensions.
The first two variants are
three-state CA models
on a square lattice.
Each site
contains either an arrow pointing upwards, an arrow pointing to the right,
or is empty.
In the first variant (Model I)
the dynamics is controlled by a traffic light, such that the right arrows
move only in even time steps and the up arrows move in odd time steps.
On even time steps each right arrow moves one step to the right
unless the site on its right hand side is occupied by another arrow
(which can be either an up or a  right arrow).
If it is blocked by another arrow it does not move, even if during
the same time step the blocking arrow moves out of that site.
Similar rules apply to the up arrows, which  move upwards.
Note that this is a fully deterministic model; randomness enters
only through the initial conditions.
In this model the traffic problem is reduced to its simplest form while
the essential features are maintained. These features include the
simultaneous flow in two perpendicular directions of objects that
cannot overlap.
No attempt is made here to draw a more direct analogy between our model
and real traffic problems.

The model is defined on a square lattice of $ N \times N $ sites
with periodic boundary conditions.
Due to the periodic boundary conditions
the total number of arrows of each type is conserved.
Moreover,
the total
number of up arrows in each column and the total number of right arrows
in each row are conserved, giving rise to $ 2N $ conservation rules.

The density of right (up) arrows is given by
$p_{\rightarrow}= n_{\rightarrow}/N^2$
($p_{\uparrow}= n_{\uparrow}/N^2$), where
$n_{\rightarrow}$
($n_{\uparrow}$)
is the number of right (up) arrows in the system.
Here we examine the case where
$p_{\rightarrow} = p_{\uparrow}=p/2$.
The (average) velocity $v$ of an arrow in a time interval
$\tau$ is defined to be the number of successful moves it makes
in $\tau$ divided by the number of attempted moves in $\tau$.
It has maximal value $v = 1$, indicating that the arrow was
never blocked, while $v=0$ means that the arrow was stuck
for the entire duration $\tau$, and never moved at all.
The average velocity
$\overline{v}$
for the system
is then obtained by averaging $v$
over all the arrows in the system.

We have performed extensive simulations of the model
starting with an ensemble of random initial conditions.
After a transient period that depends on the system size,
on $p$, and on the random initial condition,
the system reaches its asymptotic
state.
We found two qualitatively different asymptotic states, which are
separated by a sharp dynamical transition.
Below the transition all the arrows move freely in their turn and
the average velocity is $\overline{v}=1$, while above it they are all stuck
and $\overline{v}=0$ with very high probability.
A typical configuration below the transition is shown in Fig. 1,
where the system is self organized into separate rows of right and
up arrows
along the diagonals from
the upper-left to the lower-right corners.
This arrangement enables the arrows to achieve the maximal speed.
When a row of horizontal arrows moves it makes space for a row of vertical
arrows to move in the next step, such that they never collide.
Above the transition all the arrows are
stuck in a global cluster, shown in Fig. 2
(by global cluster we mean a cluster that connects one side of
the system to the other).
This global cluster is oriented along the diagonal from the
upper-right to the lower-left corners.
This way it blocks the paths of all arrows which finally get stuck
\cite{supercool}.

These two states are separated by a sharp
{\it  jamming transition}
in which the
ensemble average
velocity changes rapidly
from $\left<\overline{v}\right>=1$ to $\left<\overline{v}\right>=0$
as $p$ is varied
(see Fig. 3).
Results for five system sizes from
$16 \times 16$
to
$ 512  \times 512 $
are presented.
Small size systems
(up to $ 128  \times 128 $)
have been simulated on sequential machines,
while
the larger systems were
simulated on a DECmpp with 8k processors.
For small size systems
the transition is not sharp but there is a range
of densities
in which both asymptotically dynamic and asymptotically
static states
are found with a non-negligible probability
(depending on the initial condition).
We define $p_c(N)$ to be at the center of this region,
which is characterized by very long transients.
As the system size increases,
$p_c(N)$ tends to decrease
while the coexistence region shrinks,
giving rise to a sharper transition.
{}From our simulations we have not been able to obtain conclusive results
for $p_c$ in the infinite system limit. We find that the transition is
very sharp for large systems. However, $p_c(N)$ keeps decreasing as
$N$ increases, and we have not been able to determine whether it
converges to a finite $p_c$ or to zero in the infinite system limit.
The difficulty results from the long equilibration times
near the transition
(see Fig. 4)
and from the slow convergence of $p_c(N)$ as the system size increases.

Being a transition as a function of the concentration $p$,
between a state with no global cluster below $p_c$ to a state
with a global cluster above $p_c$, it resembles the percolation
transition
\cite{percol}.
However, the percolation transition
is a second order transition and has no dynamics.
The jamming  transition can also be considered in the
context of pinning transitions that occur in
extended systems with quenched random impurities such as
charge density waves
\cite{cdw}.
In our model there is no quenched component and
the two sets of arrows pin each other when the density increases above
threshold.

In order to examine the robustness of the jamming  transition we have also
studied a non-deterministic
variant of our model (Model II).
In Model II the traffic light is removed and
all arrows move in all time steps (unless they are stuck).
If both an up and
a right arrow try to move to the same site,
one of them will be chosen randomly,
with equal probabilities
\cite{determ}.
For this model we also find a sharp transition. The values of $p_c(N)$
are smaller than for Model I
(approximately $0.10$ for systems of size $512 \times 512$).
The value of $p_c$ in the infinite system
limit cannot be determined from our data.

By choosing a two dimensional model that has only
right and up arrows, and does not
have left and down arrows we simplify the problem without losing most of
its essential features. The essential problem that
causes traffic jams
is the need of the right and up arrows to cross each other's paths,
while each site can be occupied by only one arrow. There is no such problem
between the up and down, or between the right and left arrows, as they can
move in parallel paths that do not intersect.
In models that have both right, up, left and down
arrows one can have a stable
finite traffic jam. A simple example is a set of four arrows in which
an up arrow blocks a left arrow, which blocks a down arrow,
which blocks a right arrow, which blocks the first up arrow. This
is the {\it grid-lock} mechanism, that may occur at any density $p$.
In our model grid-locks are not possible, and the jamming transition
occurs only when a global cluster forms.

To obtain a better understanding of the model we now describe the one
dimensional analog which can be solved analytically. In one dimension there
is only one type of arrows (say right arrows) that move along
a closed ring.
Every time step each arrow moves to the right
unless it is blocked by another arrow
\cite{rule}.
The asymptotic velocity $\overline{v}$ is independent of initial
conditions.
It is
$\overline{v} = 1$
for $p < 1/2$,
while for $p>1/2$ it
decreases continuously to zero according to $\overline{v} (p) = (1-p)/p$
\cite{oned}.
We thus conclude that the sharp first order transition is indeed a result
of the interaction between the horizontal and vertical arrows due to
the excluded volume.
To clarify this point further we have performed
preliminary simulations on a variant
(Model III) in which
a  right and an up arrow are allowed to occupy the same site.
In this four-state model all arrows try to move at every time step.
If both an up arrow and a right  arrow try to move to an empty site
{\it at the same time step} they both move in and overlap.
On the other hand no arrow can move into a site which is already occupied.
This model is designed to have weaker excluded volume effects between arrows
of different types. Indeed our simulations show that model III exhibits
a continuous transition which is qualitatively similar to the one dimensional
case.

In summary, we have presented a new cellular automaton model that describes
traffic flow in two dimensions. Our simulations
of finite systems up to $512 \times 512$
show a sharp
transition that separates a low density dynamical
phase in which all cars move at a maximum speed and a
high density static phase in
which they are all stuck in a global traffic jam.
Such behavior is found both in a deterministic and a non-deterministic
variants of the model and we thus believe that it is robust and represents
a general feature of traffic flow in two dimensions.
We believe that cellular automata provide a useful framework for
traffic simulations that should be developed further.
These models are especially suitable for simulations on parallel
computers, and their flexibility is especially important in the
complex geometries of traffic networks.

This work was performed using the computational resources of the
Northeast Parallel Architectures Center (NPAC) at Syracuse University.

\figure{A typical dynamic configuration
in the low density phase below the transition.
The system is self organized into a pattern of lines
of arrows
from the upper-left to the lower-right corners
and $\overline{v}=1$.
The system size is $32 \times 32$ and $p=0.25$.}

\figure{A typical static configuration
in the high density phase above the transition.
Here the global cluster is oriented between the upper-right and
the lower-left corners, and blocks the paths of all the arrows until they
get stuck.
The system size is
$32 \times 32$ and $p = 418/1024 \approx  0.4082$.}

\figure{The ensemble average velocity $\left<\overline{v}\right>$
as a function of the concentration $p$
for five different system sizes (Model I).
As the system size increases the
transition becomes sharper, and the ensemble average velocity
changes rapidly from
$\left<\overline{v}\right>=1$ below $p_c(N)$ to
$\left<\overline{v}\right>=0$ above it.}

\figure{The median equilibration time, $T_{med}$,
for Model I as a function of $p$ for
four different system sizes.
The equilibration time is the
number of time steps it takes to reach a periodic
cycle or to get stuck.
The peak around $p_c$ becomes higher and sharper
as the system size increases up to $64 \times 64$,
and then becomes more flat for $128 \times 128$. It is not clear how to
interpret this behavior for $128 \times 128$,
although it may be that there is a very narrow
peak that we have not been able to resolve.}


\begin{references}

\bibitem{traffic}  See e.g., {\it Transportation and Traffic Theory},
Edited by N.H. Gartner and N.H.M. Wilson (Elsvier, 1987);
W. Leutzbach, {\it Introduction to the Theory of Traffic Flow}
(Springer-Verlag, 1988); D.L. Gerlough and M.J. Huber,
{\it Traffic Flow Theory} (National Research Council, Washington D.C., 1975).



\bibitem{wolf} S. Wolfram, {\it Theory and Applications of Cellular
               Automata} (World Scientific, Singapore, 1986).



\bibitem{lattgas} See e.g., {\it Lattice Gas Methods for PDE's},
                  Proceedings of the NATO Advanced Research Workshop,
                  edited by G.D. Doolen, {\it Physica} {\bf D47},
                  1-340 (1991).




\bibitem{schnitt} B. Schnittman, {\it Int. J. of Mod. Phys.}, {\bf B4},
2269 (1990).



\bibitem{bak} P. Bak, C. Tang and K. Wisenfeld, {\it Phys. Rev. Lett.},
{\bf 57}, 1111, (1987).



\bibitem{bzreac} M. Gerhardt and H. Schuster, {\it Physica} {\bf D36},
                 209 (1989).



\bibitem{supercool} Note that the fact that we
find static behavior above $p_c$ and dynamic
behavior below does not mean that there are no static configurations with
$p<p_c$ or dynamic configurations with $p>p_c$. What it means is that these
cases are atypical and have very small basins of attraction in the
ensemble of random initial conditions.
In fact the static configuration with smallest possible $p>0$ has
$p=2/N \rightarrow 0$ as $N \rightarrow \infty$, and the dynamic
configuration with $\overline{v} = 1$
and largest possible $p$ has $p=2/3$.



\bibitem{percol}  D. Stauffer, {\it Introduction to Percolation Theory}
                  (Taylor \& Francis, 1985).


\bibitem{cdw} See e.g.
    {\em Charge Density Waves in Solids}, edited by G.~Hutiray and
    J.~S\'{o}lyom (Springer-Verlag, 1985);
    G.~Gr\"{u}ner, Rev.\ Mod.\ Phys. {\bf 60}, 1129 (1988);
    {\em Charge Density Waves in Solids},
    edited by L.~P.~Gorkov and G.~Gr\"{u}ner (Elsevier, 1989).




\bibitem{determ} Note that a similar situation occurs
               in lattice gas cellular
               automata on the triangular lattice, where some states
               can have several different
               outcomes.
               In this case one can use either
               a deterministic approach such as using
               different outcomes in even and
               odd time steps or the non-deterministic approach.
               The deterministic approach is more
               efficient in numerical simulations,
               while there seems to be no significant
               difference in the results
               between the two approaches.





\bibitem{rule}   This one dimensional model is identical to Wolfram's
                 CA rule no. 184, which is asymmetric and thus does
                 not belong to the set of 32 legal rules with a
                 three site neighborhood. This model was previously
                 studied in the context of surface growth through
                 ballistic deposition (J. Krug and H. Spohn,
                 {\it Phys. Rev. } {\bf A38}, 4271 (1988)).
                 Related stochastic models
                 in two dimensions with one type of arrow
                 have also been studied
                 (S.A. Janowsky and J.L. Lebowitz,
                 {\it Phys. Rev. } {\bf A45}, 618 (1992)).




\bibitem{oned}   This result is obtained analytically by considering the
                 vacant sites as left moving arrows, with an exchange
                 dynamics such that the numbers
                 of right arrows and left arrows moving in each time
                 step are the same.






\end{references}
\end{document}